\documentclass[aps,prl,twocolumn,amsmath,amssymb]{revtex4}
\usepackage{times}
\usepackage{graphicx}
\usepackage{amsfonts}
\usepackage{amsmath, amsthm, amssymb}
\usepackage{dsfont}
\usepackage{color}

\definecolor{DarkRed}{rgb}{0.65,0,0}
\definecolor{DarkBlue}{rgb}{0,0,0.65}

\newcommand{\im}{\mathrm{i}}        
\newcommand{\ve}[1]{\boldsymbol{#1}}
\DeclareMathOperator{\diag}{diag} 

\newcommand{\nabtil}{\tilde{\nabla}}

\newcommand{\yh}{\ve{\hat{y}}}

\newcommand{\gc}{\check{g}}

\newcommand{\eg}{\textit{e.g. }}
\newcommand{\etal}{\emph{et al.}}

\begin{document}

\title{Anisotropic Paramagnetic Meissner Effect by Spin-Orbit Coupling}

\author{Camilla Espedal$^1$, Takehito Yokoyama$^2$, and Jacob Linder$^1$}

\affiliation{$^1$Department of Physics, Norwegian University of
Science and Technology, N-7491 Trondheim, Norway}

\affiliation{$^2$Department of Physics, Tokyo Institute of Technology, Tokyo 152-8551, Japan}
 
\begin{abstract}
Conventional $s$-wave superconductors repel external magnetic flux. However, a recent experiment [A. Di Bernardo \etal, Phys. Rev. X \textbf{5}, 041021 (2015)] has tailored the electromagnetic response of superconducting correlations via adjacent magnetic materials. We consider another route to alter the Meissner effect where spin-orbit interactions induce an anisotropic Meissner response that changes sign depending on the field orientation. The tunable electromagnetic response opens new paths in the utilization of hybrid systems comprised of magnets and superconductors.
\end{abstract}

\date{\today}

\maketitle
\textit{Introduction.} The Meissner effect in superconductors is 
the expulsion of magnetic fields and 
it is one of its two defining properties, the other being the absence of electrical resistance. Experiments have shown \cite{ODA1980631} that a non-superconducting material can also exhibit a Meissner response when it is in proximity to a superconductor. Via the proximity effect, superconducting correlations leak into the neighbouring metal. Intuitively, one might expect stronger superconducting correlations to give rise to a stronger Meissner response of the normal metal. The modelling of such systems via quasiclassical theory largely confirms this picture \cite{belzig1996diamagnetic}, except the observation that the magnetic susceptibility has a puzzling re-entrant behavior as a function of temperature \cite{visani1990novel,mota1996,mota_prl_00}.

When superconductors are placed in contact with ferromagnets, triplet Cooper pairs emerge that carry a net spin \cite{bergeret2005odd,buzdinrmp,linder_nphys_15, eschrig_physrep_15}. Such pairs are additionally characterized by an odd-frequency symmetry \cite{berizinskii_jetp_74} which influences several physical properties, such as the electronic density of states and electromagnetic response. Very recently, an experiment \cite{dibernardo_prx_15} observed a paramagnetic Meissner effect in an Nb/Ho/Au structure. In this system, superconductivity enhanced the magnetic signal rather than expelling it. Such a finding is of a fundamental interest, since it questions the hallmark property of perfect diamagnetism in superconductors. From a practical point of view, a paramagnetic Meissner effect  could lead to an integration of magnetic and superconducting materials in a way that has not been possible previously. Moreover, the recent demonstration of remotely induced magnetism via a superconductor reported in Ref. \cite{flokstra_nphys_15} suggests that the study of how superconductivity influences magnetic signals is particularly timely.

Motivated by these experimental advances, we show in this Letter that by combining superconductors with spin-orbit coupled materials, the Meissner effect can be modulated by the orientation of an external magnetic field. Not only does the Meissner response of the system become anisotropic as a function of field orientation, but it can even change sign. This offers a way to control the electromagnetic response of a superconducting system \textit{in situ}. In addition, we demonstrate that magnetic exchange fields $h$ that are much smaller than the superconducting gap $\Delta_0$, \eg induced via the Zeeman-effect of an external field, can lead to a similar re-entrant behavior of the susceptibility as in the experiments Ref. \cite{visani1990novel,mota1996,mota_prl_00}.

\begin{figure}[t!]
	\includegraphics[scale=1]{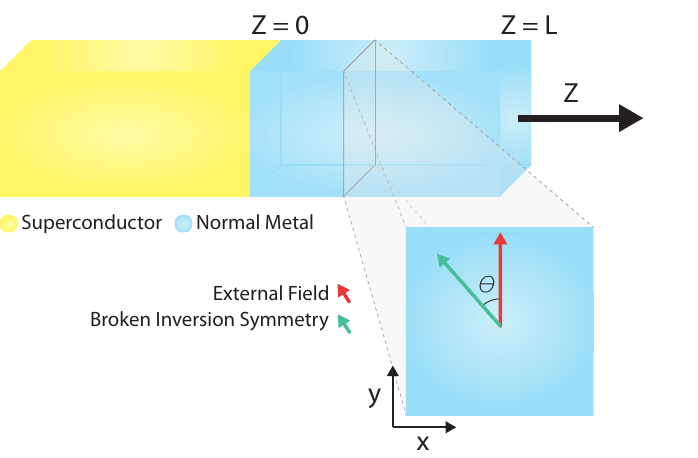}
	\caption{(Color online) The proposed experimental setup. A superconductor/normal metal (S/N) bilayer where inversion symmetry is broken in the N part, giving rise to intrinsic spin-orbit coupling. Inversion symmetry breaking is assumed throughout the N part, \eg by using a noncentrosymmetric crystal such as InAs or InSb, and the applied magnetic field is oriented an angle $\theta$ relative to the direction of the broken inversion symmetry.}
	\label{fig:system}
\end{figure}

\textit{Theory.} We consider a superconductor/normal metal (S/N) bilayer, where intrinsic spin-orbit coupling (SOC) exists in the N part. Possible candidates are materials with a noncentrosymmetric crystal structure such as InAs, which additionally has a high effective $g$-factor (strong coupling between external field and electron spins). We take into account an external magnetic field applied to this structure and describe its orientation via the angle $\theta$ (see Fig. \ref{fig:system}). The superconductor is assumed to act as a reservoir by setting its dimension much larger than the superconducting coherence length $\xi_S$. To determine the Meissner response of the system under consideration, we use the quasiclassical theory of superconductivity \cite{rammer1986quantum, serene1983quasiclassical, kopnin2001theory}. In the diffusive limit, the resulting physics is described by the Green function matrix $\hat{g}$ that solves the Usadel equation \cite{usadel1970generalized}. We use a linear-response theory \cite{narikiyo1989proximity, belzig1996diamagnetic} to compute the supercurrent due to the external magnetic field. Such an approach quantitatively accounts for experimental findings of the conventional diamagnetic Meissner effect in conventional S/N structures \cite{belzig1996diamagnetic}. Considering a strength of the external field in the range $10~\mathrm{mT}-100~\mathrm{mT}$ and assuming \cite{winkler} $g=20$ as relevant for InAs or InSb, the resulting induced Zeeman-splitting $h$ is $(0.005-0.05) \Delta_0$.  Due to the external field, a small Zeeman splitting is present in the normal metal since the Meissner response is incomplete, which we model by adding a small exchange term $h$. The SOC is accounted for by treating it as an SU(2) gauge-field \cite{bergeret2013singlet,bergeret2014spin}, included in the differentiation operator $\nabtil X = \nabla X -\im[\mathcal{\hat{A}},X]_-$, where $X$ is an arbitrary function. The inversion symmetry is broken along the unit-vector $\hat{\ve{n}} = [-\sin \theta,\cos\theta,0]$. To linear order in the momentum, this corresponds to having a Rashba-term $H_R = -\alpha(\ve{\sigma} \times \ve{k})\cdot\hat{\ve{n}}$ in the Hamiltonian where $\ve{\sigma}$ is the Pauli vector and $\hat{\ve{n}}$ points along the direction of inversion asymmetry. 
In our calculations, we fix the external field in the $y$-direction and vary $\hat{\ve{n}}$. The resulting gauge-field then reads $\hat{\mathcal{A}} = -\alpha(\sin\theta \hat{\sigma}_y + \cos\theta\hat{\sigma}_x)$. However, we emphasize that this procedure is fully equivalent to rotating the external field and keeping the sample intact which might be preferable experimentally. The S/N interface is located at $z = 0$, and the vacuum interface at $z = L$, where $L$ is the length of the N. The Usadel equation in the N is

\begin{equation}
\label{eq:usadel_equation}
	\im D \nabtil_{z} (\hat{g_N} \nabtil_{z} \hat{g_N}) = \left[\epsilon \hat{\rho}_3 + \hat{M},\hat{g_N} \right]_-,~z \in [0,L]
\end{equation}
with $\hat{\rho}_3=\text{diag}(1,1,-1,-1)$ where $D=\tau v_F^2 /3$, $\epsilon$, and $\hat{M} = h\diag(\ve{\sigma},\ve{\sigma}^*)\yh$ are the diffusion constant, the quasiparticle energy measured relative to the normal-state Fermi level, and the exchange term, respectively. The Usadel equation is accompanied by the boundary conditions \cite{kuprianov1988influence} $2 (L \Omega_N) \hat{g}^R_N \tilde{\nabla} \hat{g}^R_N = [\hat{g}^R_i,\hat{g}^R_N]_-$ at $z=0$, where $R$ indicates that we refer to the retarded component of the Green function, and $\hat{g}^R_i = \hat{g}^R_{BCS}$, where $ \hat{g}^R_{BCS}$ is the bulk BCS-solution of the Usadel equation, at the SN-interface, and $\tilde{\nabla}\hat{g}^R_N=0$ at the vacuum interface. $\Omega_N$ is a parameter describing the interface transparency. In all the numerical calculations, we use $\Omega_N = 4$. Once $\hat{g}$ has been obtained, one may compute the supercurrent density flowing through the system via the formula \cite{rammer1986quantum}:
\begin{align}
\label{eq:total_current}
\ve{j} = \frac{N_0 e D}{16} \int^\infty_{-\infty} \text{d}\varepsilon \text{Tr}\{ \hat{\rho}_3(\gc \tilde{\nabla}'\gc)^\text{K} \}.
\end{align}
where the covariant derivative, $\tilde{\nabla}'$, contains both the U(1) electromagnetic vector-field and the SU(2) SOC-field. The Green function matrix,  $\check{g}$,  includes the retarded, advanced, and Keldysh components \cite{kopnin2001theory}. We find the Meissner response current from Eq. (\ref{eq:total_current}), by extracting the term which is proportional to the electromagnetic vector potential,

\begin{equation}
\label{eq:diamagnetic_current}
	j^d_i = -\im \frac{N_0 e^2 D}{16}A_i(\ve{r})\int^\infty_{-\infty} \text{d}\epsilon~ j_{\epsilon,i}^d
(\ve{r},\varepsilon)\tanh\left(\frac{\beta \epsilon}{2}\right)
\end{equation}
where $ j_\epsilon^d (\ve{r},\epsilon)=\text{Tr}\{  (\hat{\rho}_3 \hat{g}^R)^2 - (\hat{\rho}_3 \hat{g}^A)^2\}$. In linear response, we solve the Usadel equation without the electromagnetic vector potential and use the solution for $\hat{g}$ in Eq. (\ref{eq:diamagnetic_current}) in order to find the Meissner current. One can show that the contributions from $\nabla$ and $\hat{\mathcal{A}}$ in $\tilde{\nabla}'$ to the current in Eq. (\ref{eq:total_current}) vanish.  The last step consists of solving the Maxwell equation $\nabla \times \boldsymbol{B} = \mu_0 \ve{j}$ in order to obtain the magnetic vector potential $\boldsymbol{A}(z)$ in the normal metal. Having determined $\boldsymbol{A}(z)$, we may then compute \eg the local supercurrent or the magnetic susceptibility which both probe the Meissner-response of the system. We choose the London gauge $\ve{A} = [A(z),0,0]$, and obtain
\begin{equation}
	\partial_z^2 A(z) = -\mu_0 j^d_x 
\end{equation}
We assume that the applied field is completely screened within the bulk superconductor, and that there is no screening at the vacuum edge of N \cite{narikiyo1989proximity}. With these assumptions, the boundary conditions become $A(z=0) = 0$, $\frac{\text{d} A}{\text{d} z}(z=L)	= \mu_0 H$. The susceptibility, $\chi$, the response of the material to the external field integrated over $L$, is then obtained as:
\begin{equation}
	\chi = A(L)/\mu_0 HL - 1.
\end{equation}

Having solved the Maxwell equation, one can also find the magnetization, which is related to the vector potential $\ve{M}(z) = (1/\mu_0)\ve{B}(z) - H$. For our system, having $\ve{B} = \nabla \times \ve{A}(z)$ pointing in the $y$-direction, we get $\ve{M}(z) = (0,M(z),0)$. In our simulations, we set $N_0e^2D\Delta_0\mu_0L^2/(16\hbar) \equiv k = 16$ and $\xi_S/L=0.3$. We have verified that altering the value of $k$ in a wide range of five orders of magnitude does not influence the results qualitatively, and hence the results presented herein are representative. 

We have solved the above set of differential equations (the Usadel and Maxwell equations) numerically, utilizing the Ricatti-parametrization  \cite{eschrig2004singlet} for the quasiclassical Green function $\hat{g}$ extended to include SOC \cite{jacobsen2015critical, jacobsen_prb_15b}.

\textit{Results}. Before providing a fully numerical solution, it is instructive to consider how the exchange field and SOC induce triplet Cooper pairs. The odd-frequency symmetry of these pairs results in a paramagnetic Meissner response in contrast to the conventional spin-singlet pairs which generate a screening supercurrent \cite{yokoyama_prl_11, mironov_prl_12, alidoust_prb_14, fominov_prb_15}. Assuming a weak proximity effect, one derives the following diffusion equations for the singlet $f_s$ and triplet $\ve{f}_t$ superconducting correlations:
\begin{align}
\label{eq:lin_usadel}
	\frac{\im}{2}D \partial_z^2 f_s(z) &= \epsilon f_s (z) -\ve{h} \cdot \ve{f}_t(z),\notag\\
	\frac{\im}{2}D \partial_z^2 \ve{f}_t (z) = & (\epsilon + \mathcal{R}) \ve{f}_t(z) - \ve{h} f_s (z), 
\end{align}
where the exchange field points in the $y$-direction so that $\ve{h} = h\yh$, $h$ is the strength of the exchange field, and $\mathcal{R}=2\im D \big(\alpha^2\ve{\Omega}_r + \alpha\ve{\Omega}_p \partial_z \big)$. We define $s\equiv\sin\theta,\; c\equiv\cos\theta$, and:
	\begin{equation}
		\label{eq:coupling_matrices}
	\ve{\Omega}_r = 
		\begin{pmatrix}
			-s^2 & cs & 0 \\
			-cs & c^2 & 0 \\
			0 & 0 & 1
		\end{pmatrix}
		,\; \ve{\Omega}_p = 
		\begin{pmatrix}
			0 & 0 & s \\
			0 & 0 & -c \\
			s & c & 0
		\end{pmatrix}.
	\end{equation}
	
	\begin{figure}[t!]
\includegraphics[scale=0.55]{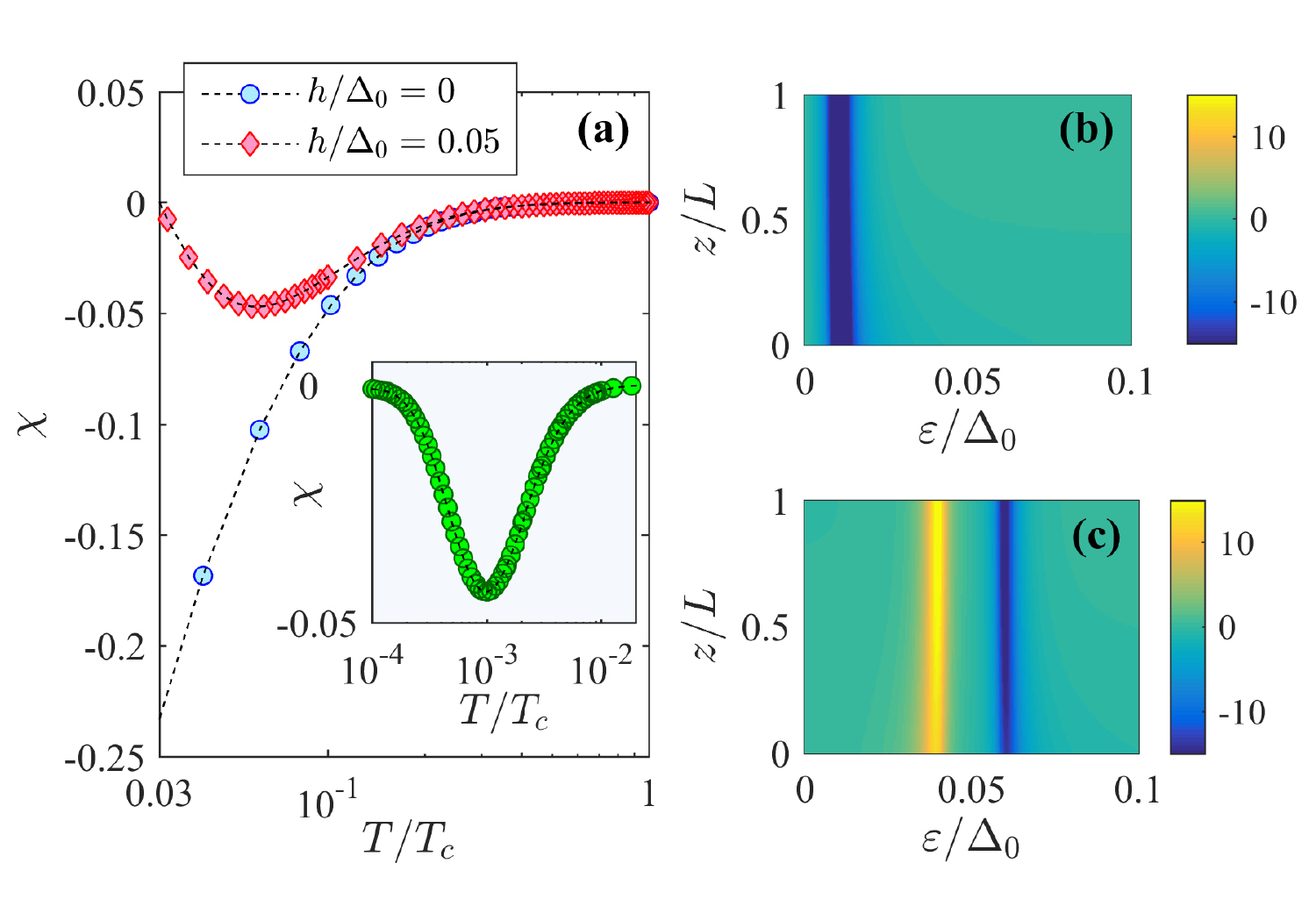}
\caption{(Color online) (a) Magnetic susceptibility $\chi$ vs. temperature 
in the absence of SOC, $\alpha=0$. When a small exchange field is present, the re-entrant effect comes into play. The dashed lines are added as a guide to the eye. \textit{Inset:} Re-entrance effect for $\xi_S/L=0.043$ and $h = 0.001 \Delta_0$. Spectral Meissner response, $-\im j^d_{\epsilon}$, for (b) $h/\Delta_0 = 0$ and (c) $h/\Delta_0 = 0.05$ shows that a positive contribution to the spectral current appears when we add the exchange field. Note while we only plotted for positive energy, the spectral current is an odd function of the energy, while $-\im j^d_{\epsilon} \tanh{\beta\epsilon/2}$ is even.}
\label{fig:chiandcurrent}
\end{figure}
	
The exchange field $h$ induces the triplet component $\ve{f}_t \parallel \boldsymbol{h}$. The SOC, however, allows the coupling to the triplet Cooper pairs to be varied by changing $\theta$, i.e. rotating the magnetic field. As will be demonstrated below, the fact that the Meissner response becomes highly anisotropic with regard to variations in $\theta$ indicates that the triplet generation is an important factor in determining the electromagnetic response of the superconducting correlations. In particular, one should note that for an orientation $\theta = \pi/2$ of the external field, we can see from Eqs. (\ref{eq:lin_usadel}) and  (\ref{eq:coupling_matrices}) that the triplet components become decoupled and the situation is equivalent to having no SOC. Our numerical results are also consistent with this statement.

We consider first the case \textit{without} SOC for a long normal metal as in the experiments of Refs. \cite{visani1990novel,mota1996, mota_prl_00}. The magnetic susceptibility as a function of temperature $T$ is shown in Fig. \ref{fig:chiandcurrent}(a) both with and without an exchange field. For $h=0$, we see as expected a conventional diamagnetic Meissner effect. Interestingly, even for a very small exchange field $h/\Delta_0=0.05$ we observe that $\chi$ vs. $T$ displays a re-entrant behavior. To understand the physical origin of this behavior, we first note that when the normal metal length satisfies $L\gg\xi_S$, the proximity-induced minigap is much smaller than the superconducting gap $\Delta_0$. The minigap is determined by the Thouless energy \cite{pilgram2000excitation} $\varepsilon_T = D/L^2$, so that longer samples have smaller minigaps. Now, in the presence of exchange fields $h/\Delta_0 \sim \varepsilon_T / \Delta_0 = (\xi_S/L)^2$, the triplet proximity effect becomes resonant and results in a zero-energy peak in the density of states \cite{yokoyama2007manifestation}. For smaller minigaps (larger $L$), only a small exchange field is needed to get sufficiently close to resonance, which explains why a long normal metal can be influenced by a small $h$. As seen in the inset of Fig. \ref{fig:chiandcurrent} (a), a change in $\xi_S/L$ will also give re-entrant behaviour as long as $h$ is lowered accordingly. Since $\varepsilon_T$ is now smaller, the minimum of $\chi$ occurs at a lower temperature.
 
The resonant behavior also influences the spectral current density and inspection of this quantity reveals why the Meissner current, and in turn susceptibility $\chi$, behaves non-monotonically. Consider Fig. \ref{fig:chiandcurrent}(b) and (c) where we have plotted the spectral current without and with the exchange field. When the exchange field is turned on, there is both a positive and negative contribution in the spectral supercurrent. What ultimately determines the total Meissner supercurrent in Eq. (\ref{eq:diamagnetic_current}) is how these low-energy contributions are weighted by the distribution function factor $\tanh(\beta \epsilon /2)$. At low temperatures, where $\beta$ is large, the positive contribution to the shielding supercurrent is weighted more efficiently. Increasing the temperature shifts the weight toward the negative contribution and the Meissner response becomes more diamagnetic. There thus exists a crossover temperature regime where there is a competition between these two phenomena, which leads to the re-entrant effect shown in Fig. \ref{fig:chiandcurrent}(a). 

\begin{figure}[t!]
\includegraphics[scale=0.49]{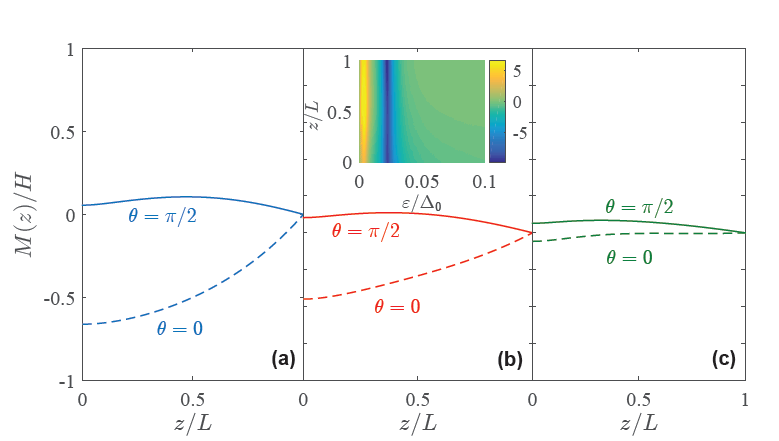}
\caption{(Color online) The magnetization profile, $M(z)$, as a function of position inside N for $T/T_c=10^{-3}$, $\alpha\xi_S = 0.75$ and (a) $h/\Delta_0 = 0.025$, (b) 0.05, and (c) 0.1. In each case, we show two different magnetic field orientations, $\theta = \pi/2$ and $\theta = 0$, which demonstrate the switching between a diamagnetic and paramagnetic Meissner responses. We underline that the positive magnetization $M(z)$ shown in panel (b) is consistent with Fig. \ref{fig:chiandcurrent}(a) because $\chi$ is positive at very low temperatures $T/Tc<0.03$ when $h/\Delta_0=0.05$. \textit{Inset}: the spectral supercurrent for $\theta=0$, demonstrating how triplet pairing is present (positive current near $\varepsilon/\Delta_0\simeq 0.005$) whereas the diamagnetic contribution is larger (negative current near $\varepsilon/\Delta_0\simeq 0.02$). }
\label{fig:switching}
\end{figure}

We now turn to the effect of \textit{including} SOC and show that altering the field orientation $\theta$ causes a transition from a standard Meissner effect to a paramagnetic Meissner response. This pertains uniquely to the presence of SOC: in its absence, the Meissner response is completely independent of the field orientation. To see that this is a robust effect that occurs over a broad range of exchange field values, we have plotted in Fig. \ref{fig:switching} the supercurrent-induced magnetization response for $\theta=0$ and $\theta=\pi/2$ for several values of $h$. It is found that rotating the field by 90$^\circ$ (from $\theta=0$ to $\theta=\pi/2$) now inverts the sign of the orbital response and the supercurrent generates a magnetization that enhances the net magnetic field. To the best of our knowledge, this is the first prediction of how the Meissner response in a normal metal can be inverted \textit{in situ}. 

The physical origin of this phenomenon can be traced back to how the generation of triplet Cooper pairs depends on the magnetic field orientation $\theta$, as seen from the analytical equations. For $\theta=0$, Eqs. (\ref{eq:lin_usadel}) and (\ref{eq:coupling_matrices}) show that the various triplet components of $\ve{f}_t$ are coupled, which might suggest that the Meissner response should become more paramagnetic due to the increased pathways to create triplets. In contrast, we see in Fig. \ref{fig:switching} that the opposite takes place: a diamagnetic effect occurs for $\theta=0$, while the signal becomes paramagnetic for $\theta=\pi/2$ (where the SOC has no effect). The reason for this is that besides coupling the components of $\ve{f}_t$, the SOC has an additional consequence: it introduces a depairing effect on the triplet correlations in the system due to the term $\propto \boldsymbol{\Omega}_r$ which adds an imaginary component to the quasiparticle energy. We find this effect that typically dominates for weak exchange fields and hence it restores the orbital response of the system to a conventional Meissner effect. The modification of the spectral supercurrent due to the presence of SOC for $\theta=0$ is shown in the inset of Fig. \ref{fig:switching}: the presence of triplets is manifested via a positive contribution to the current, whereas the diamagnetic response (negative peak) is larger and results in a net conventional Meissner response.

\textit{Discussion}. Previously, paramagnetic Meissner effects have been discussed in the context of high-$T_c$ superconductors \cite{kostic_prb_96,higashitani1997mechanism, shan_prb_05,anlage_prl_13}. For such a system, the presence of Andreev surface-bound states give a paramagnetic contribution to the shielding supercurrent but are not strong enough to render the total Meissner response paramagnetic in large superconductors \cite{suzuki2014paramagnetic}. Ref. \cite{fauchere_prl_99} showed that repulsive interactions in the N can induce a midgap bound state at an SN interface, leading to a paramagnetic Meissner effect. It is also important to emphasize that metastable paramagnetic Meissner effects have been shown to originate from other types of effects which are not related to unconventional superconductivity, but instead to flux capturing at the surface for small superconductors \cite{geim_nature_98}. In this case, the conventional Meissner state is restored by external noise. This scenario are distinct from that of the present paper, where the paramagnetic Meissner effect occurs due to an exotic type of odd-frequency superconductivity. 

The effects predicted in this work require a local magnetization probe. This could be accomplished using low-energy muon spin-spectroscopy which offers a very high sensitivity to magnetic fields ($<0.1$ G) \cite{dibernardo_prx_15}. Alternatively, one could use a nano-SQUID technique which is known to feature single-spin sensitivity \cite{vasyukov_naturenano_13}. Finally, we note that odd-frequency triplet pairing has recently been predicted to occur in S/N systems with SOC \cite{reeg_prb_15} where the case of strong SOC was included. It would be of interest to study the Meissner effect in this regime, which goes beyond the quasiclassical approximation.

\textit{Conclusion}. Summarizing, we have shown that SOC in the normal metal fundamentally alters its Meissner response when placed in proximity to a superconductor. The supercurrent-induced magnetization displays anisotropic behavior depending on the orientation of the applied field, and can even switch the sign. This provides a way to control the electromagnetic response of superconducting structures, swapping between a conventional and an inverse Meissner response. In addition, we have shown that a re-entrant effect of the magnetic susceptibility can occur in S/N-structures in the presence of very small exchange fields $h\ll\Delta_0$. From our simulations, we find that for such fields, triplet pairing can play an important role in determining the magnetic properties of a material when its length greatly exceeds $\xi_S$.

\textit{Acknowledgments}. We thank A. Di Bernardo, A. Brataas, J. A. Ouassou, and S. Jacobsen for useful discussions. Funding 
via the “Outstanding Academic Fellows” programme at NTNU, the
COST Action MP-1201 and the Research Council of Norway Grant
numbers 205591, 216700, and 240806, is gratefully acknowledged.

\bibliography{bibliography}

\end{document}